\documentclass[prb,showpacs,twocolumn,aps]{revtex4}
\usepackage{graphicx,rotating,epsfig}
\newcommand{\be}{\begin{equation}}
\newcommand{\ee}{\end{equation}}
\newcommand{\bea}{\begin{eqnarray}}
\newcommand{\eea}{\end{eqnarray}}
\newcommand{\up}{\uparrow}
\newcommand{\down}{\downarrow}
\newcommand{\bwt}{\begin{widetext}}
\newcommand{\ewt}{\end{widetext}}
\newcommand{\ham}{\mathcal{H}}
\begin{document}
\title{
Proximity effect in clean strong/weak/strong superconducting tri-layers
}
\author{Lucian Covaci and Frank Marsiglio}
\affiliation{
Department of Physics, University of Alberta, Edmonton, Alberta,
Canada, T6G~2J1}
\begin{abstract}

Recent measurements of the Josephson critical current through
LSCO/LCO/LSCO thin films showed an unusually large proximity effect.
Using the Bogoliubov-de Gennes (BdG) equations for a tight binding
Hamiltonian we describe the proximity effect in weak links between a
superconductor with critical temperature $T_c$ and one with critical
temperature $T_c$', where $T_c>T_c$'. The weak link (N') is
therefore a superconductor above its own critical temperature and
the superconducting regions are considered to have either s-wave or
d-wave symmetry. We note that the proximity effect is enhanced due
to the presence of superconducting correlations in the weak link.
The dc  Josephson current is calculated, and we obtain a non-zero
value for temperatures greater than $T_c$' for sizes of the weak
links that can be almost an order of magnitude greater than the
conventional coherence length. Considering pockets of
superconductivity in the N' layer, we show that this can lead to an
even larger effect on the Josephson critical current by effectively
shortening the weak link.

\end{abstract}

\date{\today}
\maketitle
\section{Introduction}

The proximity effect between a superconductor and a normal metal has
been thoroughly investigated using various techniques:
Ginzburg-Landau theory \cite{ginzburg,chen}, quasi-classical Green function
methods \cite{kogan}, Gorkov equation methods \cite{gennes64,macmillan,werthamer,wu}, and tight-binding BdG
methods \cite{hirsch,zhu,halterman}. From an experimental point of view, one of the
better suited experiments is the measurement of the Josephson
critical current in weak links \cite{likharev}.

In a recent experiment \cite{bozovic} an unusually large proximity
effect is reported, and the authors argue that it cannot be
explained by the conventional proximity effect. The system used in
the experiment is a c-axis oriented one. The c-axis Josephson
critical current is measured through a thin film system made of
doped LCO with $T_c=25$K, sandwiched between optimally doped LSCO
with $T_c=45$K. The thin film is considered to be in the clean limit
and because of the epitaxial growth of the films the transmission at
the interfaces is close to unity and interface roughness is on the
order of the lattice constant. In a particular setup, the LCO thin
film used had a thickness of $100$\AA{}. Fitting the critical
current around $T_c^\prime$ the authors extract a coherence length
in the LCO film which is two orders of magnitude larger than
expected. Because of this discrepancy and the observation of
non-zero critical current for $T<30$K the authors reported this
effect to be a ``giant proximity effect".

Although the Josephson junction has been thoroughly investigated in
the past for both s-wave \cite{golubov} and d-wave symmetries
\cite{tanaka1,tanaka2,tanaka3,tanaka4,tanaka5,delin}, we feel that
the calculation of the Cooper pair leaking distance in the case of
clean limit and superconducting weak links needs further
investigation. We are interested to observe if the leaking distance
will be influenced by the finite critical temperature of the weak
superconductor.

We propose the use of the numerical solutions of the BdG equations
in a tight binding formulation in order to obtain a direct
calculation of the coherence length and of the Josephson critical
current. In the clean limit the BdG equations are particularly easy
to solve because impurity averaging is not required. This method is
complementary to the quasi-classical methods used in the dirty
limit, namely the Usadel equations \cite{usadel,cuevas05}.

For coherent transport in the c-axis direction, the properties of
the Josephson current for d-wave superconductors will be similar to
the properties of the current for s-wave superconductors. For planar
interfaces, with $\hat{z}$ the direction perpendicular to the
interfaces, the d-wave order parameter will have no $k_z$
dependence, $\Delta(k_x,k_y,k_z) \sim \Delta_0
(cos(k_xa)-cos(k_ya))$ and therefore will have properties similar to
a superconductor with s-wave symmetry. When Fourier transforming the
$\hat{x}$ and $\hat{y}$ directions, and considering an effective 1D
problem in the $\hat{z}$ direction, the d-wave order parameter will
be due to an effective on-site interaction within each ab-plane. We
will calculate the Josephson current in the c-axis direction for a
3D d-wave superconductor. We will also show calculations of the
Josephson critical current and the Cooper pair leaking distance for
a 2D s-wave superconductor and for the 100 interface of a 2D d-wave
superconductor.

The giant proximity effect is observed in underdoped cuprates, for
temperatures $T>T_c^\prime$ for which the middle layer is considered
to be in the pseudo-gap state. Previous theoretical investigations
of the giant proximity effect \cite{kresin,dagotto} considered the
N' layer to be comprised of pockets of superconductivity. In a recent theoretical study \cite{zhu05}, interstitial oxygen dopants are considered to modify locally the pairing interactions. The disordered dopants are enhancing the pairing interactions, thus increasing the size of the local gap. This was observed in recent STM experiments \cite{mcelroy} in BSCCO, which showed that the regions of enhanced superconductivity are correlated with the positions of the interstitial oxygen atoms.  Because of
the proximity effect, the superconducting pockets will be coupled
and current will flow through percolating paths. The presence of
these pockets will effectively shorten the length of the weak link
and the strong external superconductors will be coupled for values
of the effective length comparable with the leaking distance. The
modification of the leaking distance due to the finite value of
$T_c^\prime$ will have an important influence on the effective
length. Considering equally spaced areas of strong superconductivity
with critical temperature $T_c$ embedded in the weak superconductor
with $T_c^\prime$ we will calculate the critical Josephson current
and find its dependence on the length of the weak link and on the
volume of the embedded superconducting pockets. If one considers
disordered regions of strong superconductivity in the $a-b$ planes
of a high-T$_c$ superconductor, then the distance between two
pockets from different $Cu-O$ planes will be normally distributed.
The equally spaced pockets scenario should be the one that gives
maximal Josephson current and will give insight about the influence
of these pockets on the current.

This paper is organized as follows. In the next section we will
present our method. While the BdG procedure on a lattice is now well
known, we nonetheless include some details, as some ``standard"
approximations are included for clarity. The treatment of infinite
surfaces will be outlined. In the third section we apply the BdG
equations to a tri-layer system and show results of the calculation
of the order parameter, the leaking distance and the dc Josephson
current. Both the cases of s-wave and d-wave symmetries of the
superconducting order parameters are considered.
We find that the proximity effect can be considerably enhanced at
temperatures close to (but above) the critical temperature of the
weak superconductor. The presence of randomly distributed pockets of superconductivity in N' enhances dramatically the Josephson critical current and leads to a ``giant proximity effect''.

\section{Method}

In order to describe the superconducting state we use the tight
binding extended Hubbard Hamiltonian:

\begin{eqnarray}
\ham &=& -\sum_{<ij>\sigma}t_{ij}c^{\dagger}_{i\sigma}c_{j\sigma}
-\mu \sum_{i\sigma}c^{\dagger}_{i\sigma}c_{i\sigma} \nonumber \\
&+&\sum_i
U_i n_{i\uparrow} n_{i\downarrow}+\frac{1}{2}\sum_{<ij>\alpha
\beta}V_{ij}n_{i\alpha}n_{j\beta},
\end{eqnarray}
where $t_{ij}$ is the nearest neighbor hopping amplitude which describes the kinetic
energy, $\mu$ is the chemical potential used to fix the filling of
the system, $U_i$ is the on-site interaction, $V_{ij}$ is the
nearest neighbor interaction and $n_{i\sigma}=c_{i\sigma}^\dagger
c_{i\sigma}$ is the density operator at site $i$ corresponding to spin $\sigma$.

The properties of this Hamiltonian have been studied previously
\cite{micnas}; it should be viewed as an effective Hamiltonian with
which one can describe s-wave and d-wave symmetries of the
superconducting order parameter. For an s-wave superconductor we
choose an attractive on-site interaction $U_i<0$ and no
nearest-neighbor interaction $V_{ij}=0$, while for a d-wave
superconductor we set the nearest neighbor interaction to be
attractive $V_{ij}<0$ and the on-site interaction to vanish or be repulsive.
The interaction parameters $U_i$ and $V_{ij}$ are dependent on position,
breaking translational invariance. This will allow us to
describe interfaces between different types of materials.

Using the Hartree-Fock mean field decomposition this Hamiltonian can
be transformed into a one-particle mean-field Hamiltonian:

\begin{eqnarray}
\ham & = & \sum_{<ij>\sigma} (-t_{ij}-\delta_{ij} \mu) c_{i\sigma}^\dagger c_{j\sigma} + \sum_{i} (\Delta_i c_{i\up}^\dagger c_{i\down}^\dagger + h.c.) \nonumber \\
& + & \sum_{<ij>} (\Delta_{ij} (c_{i\up}^\dagger c_{j\down}^\dagger + c_{i\down}^\dagger c_{j\up}^\dagger) + h.c.).
\end{eqnarray}

For planar junctions, infinite surfaces can be considered and
therefore translational invariance in the direction parallel to the
surface is recovered. By doing a Fourier transform of the
Hamiltonian in the direction parallel to the surface we only have to
solve one-dimensional inhomogeneous problems. For any point in
k-space the problem becomes a one-dimensional inhomogeneous problem. In the case of a 2D superconductor with an infinite surface along the $\hat{y}$ direction the Hamiltonian becomes:
\begin{eqnarray}
\ham &=& \sum_{k_y}\sum_{<ij>\sigma}
-(1-\delta_{ij})t_{ij}^{\perp}-\delta_{ij}[\mu+2t_i^{||}cos(k_ya))]
c_{i\sigma}^{k_y\dagger} c_{j\sigma}^{k_y} \nonumber\\ &+&
\sum_{k_y}\sum_{i}
[\Delta_i+2\Delta_{i}^{||}cos(k_ya)]c_{i\up}^{k_y\dagger}
c_{i\down}^{k_y\dagger} + h.c. \nonumber\\ &+&
\sum_{k_y}\sum_{<ij>} \Delta_{ij}^{\perp} [c_{i\up}^{k_y\dagger}
c_{j\down}^{k_y\dagger} + c_{i\down}^{k_y\dagger}
c_{j\up}^{k_y\dagger}] + h.c.,
\end{eqnarray}
where $i$ and $j$ are now in the direction perpendicular to the
surface and $a$ is the lattice constant. $t^\perp$ and $\Delta_{ij}^\perp$ are the hopping amplitude and pair potential in the direction perpendicular to the surface and $t^{||}$ and $\Delta_i^{||}$ are the hopping amplitude and the pair potential in the direction parallel to the surface. The mean-field order parameters are to be calculated self-consistently:
\begin{eqnarray}
\Delta_{i} &=& \frac{U_i}{N_y} \sum_{k_y}  <c_{i\down}^{k_y}c_{i\up}^{k_y}>, \\
\Delta_{i}^{||} &=& \frac{V_{i}^{||}}{N_y} \sum_{k_y}  <c_{i\down}^{k_y}c_{i\up}^{k_y}> cos(k_ya), \\
\Delta_{ij}^{\perp} &=& \frac{V_{ij}^{\perp}}{2N_y} \sum_{k_y}
(<c_{i\down}^{k_y} c_{j\up}^{k_y}> +
<c_{i\up}^{k_y} c_{j\down}^{k_y}>),
\end{eqnarray}
where $\Delta_{i}$ is the s-wave order parameter, $\Delta_{i}^{||}$ is the d-wave order parameter of a link in the direction parallel to the surface and $\Delta_{ij}^{\perp}$ is the d-wave order parameter of a link in the direction parallel to the surface.

In the 3D c-axis geometry, the surface is considered to be in the $\hat{x}-\hat{y}$ plane. After Fourier transforming in these directions, the Hamiltonian becomes:
\begin{widetext}
\begin{eqnarray}
\ham &=& \sum_{k_x k_y} \sum_{<ij>\sigma}
-(1-\delta_{ij})t_{ij}^{\perp}-\delta_{ij}\{\mu+2t_i^{||}[cos(k_xa)+cos(k_ya)]\}
c_{i\sigma}^{k_x k_y \dagger} c_{j\sigma}^{k_x k_y} \nonumber\\
&+& \sum_{k_x k_y} \sum_{i}
\{\Delta_i+2\Delta_{i}^{||}[cos(k_xa)-cos(k_ya)]\}c_{i\up}^{k_x k_y \dagger}
c_{i\down}^{k_x k_y \dagger} + h.c..
\end{eqnarray}
\end{widetext}
The self-consistency in the order parameters is now given by the following equations:
\begin{eqnarray}
\Delta_{i} &=& \frac{U_i}{N_xN_y} \sum_{k_x,k_y}  <c_{i\down}^{k_x k_y}c_{i\up}^{k_x k_y}>,\\
\Delta_{i}^{||} &=& \frac{ V_{i}^{||}}{N_xN_y}\sum_{k_x,k_y} <c_{i\down}^{k_x k_y}c_{i\up}^{k_x k_y}> [cos(k_xa)-cos(k_ya)] \,\,\,\,\,\,\,\,\,
\end{eqnarray}
where $i$ is now taken to be the site index in the $\hat{z}$
direction. $\Delta_i$ is the on-site s-wave order parameter while
$\Delta_i^{||}$ is the d-wave order parameter which has components
only in the $\hat{x}$ and $\hat{y}$ directions.

We follow the standard procedure \cite{gennes} of introducing a
canonical transformation of the electron operators:
\begin{eqnarray}
c_{i\up}^{k_x k_y}&=&\sum_n u_n^{i k_x k_y} \gamma_{n\up} + v_n^{i k_x k_y \ast} \gamma_{n\down}^\dagger \\
c_{i\down}^{k_x k_y}&=&\sum_n u_n^{i k_x k_y} \gamma_{n\down} - v_n^{i k_x k_y \ast} \gamma_{n\up}^\dagger.
\end{eqnarray}
This transformation will diagonalize the Hamiltonian and one obtains
the BdG equations for each pair of momentum vectors $k_x$, $k_y$:
\begin{equation}
\left(
\begin{tabular}{lcr}
$H^{0 k_x k_y}$ & $\Delta^{k_x k_y}$\\
$\Delta^{k_x k_y \ast}$ & $-H^{0 k_x k_y}$\\
\end{tabular}
\right) \left(\begin{tabular}{lcr} $u^{k_x k_y}$\\ $v^{k_x k_y}$ \end{tabular}\right)=
\epsilon ^{k_x k_y}\left( \begin{tabular}{lcr} $u^{k_x k_y}$\\ $v^{k_x k_y}$ \end{tabular}\right).
\end{equation}
These equations describe the quasi-particle states in inhomogeneous
superconductors. The BdG equations are equivalent to an eigenvalue
problem with parameters that require self-consistent calculation. We start with
an initial guess for the order parameter profile
and we diagonalize the resulting
Hamiltonian. In our infinite-surface setup, we need to diagonalize
a one-dimensional Hamiltonian for every point in momentum space. Using
the self-consistency equations (4)-(6) we recalculate the order
parameter profile. The solution is obtained when the difference in
the order parameters between two steps is smaller than a desired
accuracy.

The self-consistent calculation of the order parameter ensures that
the order parameter in the ``normal metal" region has knowledge
about the pair potential in this layer. If the initial guess is a
step function, i.e. the order parameter in the middle region is
zero, after one iteration the pair amplitude will become non-zero
because of the proximity effect. In the case $U^\prime = 0$, the
order parameter will remain zero: $\Delta \sim U_i^\prime
<c_{i\up}c_{i\down}>$, while for the case $U^\prime < 0$ the new
order parameter has a finite value throughout the layer. If we were
to fix the order parameter in the superconducting regions (we do
not), the $U^\prime=0$ solution would need only one iteration to
converge.

The BdG formalism allows us to calculate the dc current in the
absence of applied voltages. In the tight-binding formulation the
current operator is:
\begin{equation}
J_{ij}=\sum_\sigma t_{ij}(c_{i\sigma}^\dagger c_{j\sigma} - c_{j\sigma}^\dagger c_{i\sigma}).
\end{equation}
The expectation value of the current will be non-zero only if the order parameters in the two superconducting layers
have different phases. For the mean-field Hamiltonian with s-wave order parameters one gets for the continuity
equation:
\begin{eqnarray}
<\frac{\partial n_i}{\partial t}>&=&<[H,n_i]> \nonumber\\
&=&
\sum_\sigma t_{ij}(<c_{i\sigma}^\dagger c_{j\sigma}> - <c_{j\sigma}^\dagger
c_{i\sigma}>) \nonumber \\
&+& <c_{i\down}c_{i\up}>\Delta_i^\star -
<c_{i\up}^\dagger c_{i\down}^\dagger> \Delta_i
\end{eqnarray}

If the order parameter is calculated self-consistently $\Delta_i
\sim <c_{i\down}c_{i\up}>$, then we recover the continuity equation,
$<\frac{\partial n_i}{\partial t}>=<J_{ij}>$. Otherwise, if the order parameters are not calculated self-consistently but set to a desired value (the case of hard boundary) then the last two terms in Eq. (14) can be seen as current source terms.

In our calculation the coherence factors, $u_k$ and $v_k$, are
complex numbers and they will give the magnitude and the phase of the order
parameters. The magnitude of the order parameters is calculated self-consistently and after each iteration the phase of the external layers is set to a desired value. For the calculation of the dc Josephson current we restrict the
phase of the two superconductors to a desired phase difference,
while for the weak link, the phase is calculated self-consistently.

\section{Results for SN$^\prime$S}

For the $SN^\prime S$ tri-layers the interactions are only on-site
attractive interactions. The value of the parameter $U_i$ will set
the magnitude of the order parameter throughout the sample. We
consider the following setup (Fig. 1), $U_i=U$ for $0<i<A$ and
$B<i<N$, while $U_i=U^\prime$ for $A<i<B$. In this particular case
$V_{ij}$ is vanishing, because we ignore the d-wave symmetry. The
value of $U^\prime$ is chosen so that $|U^\prime|<|U|$, allowing us to describe
the $N^\prime$ material with a lower critical temperature $T^\prime_c$. For
temperatures greater than $T^\prime_c$ the $a-b$ region cannot sustain
superconductivity by itself. The order parameter will leak from the
stronger superconductors, and the characteristic length is called
the ``leaking" distance.

\begin{figure}[ht]
\begin{center}
\begin{turn}{0}
\epsfig{figure=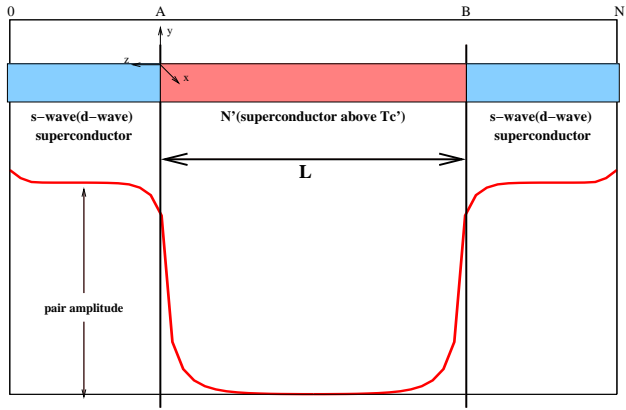,width=\columnwidth}
\end{turn}
\caption{(Color online) The pair amplitude profile through a
Josephson junction. The pair amplitude is shown for $U=-3t$ and
$U^\prime=0$. The regions $0-A$ and $B-N$ are superconducting, while
the region $A-B$ is a superconductor above its critical temperature.
The interfaces are considered to have perfect transmission and the
whole system is in the clean limit.}
\end{center}
\end{figure}

Fig.1 shows the order parameter profile for $U=-3t$, $U^\prime=-2t$ and
$T=0.21t$. Note that $T^\prime_c = 0.205t$ for the weak superconductor,
while $T_c = 0.459t$ for the strong one.
Similar to the $T^\prime_c=0$ case the order parameter has an
exponential dependence on distance away from the interface,
$\Delta \sim \Delta_0 exp(-x/\xi)$. This
is true only for temperatures much larger than $T^\prime_c$ and for distances
from the interface greater than the coherence length of the stronger
superconductors. The coefficient of the exponential decay is given
by the leaking distance, $\xi$. In the normal metal case ($T_c^\prime = 0$ K) the
clean limit leaking distance is inverse proportional to the
temperature:
\begin{equation}
\xi=\frac{\hbar v_F}{k_B T}.
\end{equation}

For the case of a weak superconductor, the relevant temperature scale
is $T - T_c^\prime$.
Plotting the order parameter versus distance from the interface on a
semi-log scale (Fig. 2) for different temperatures, we can extract
the leaking distance. As expected from the $T_c^\prime = 0$ K case, the
leaking distance is decreasing with increasing temperature.

\begin{figure}[ht]
\begin{center}
\begin{turn}{-90}
\epsfig{figure=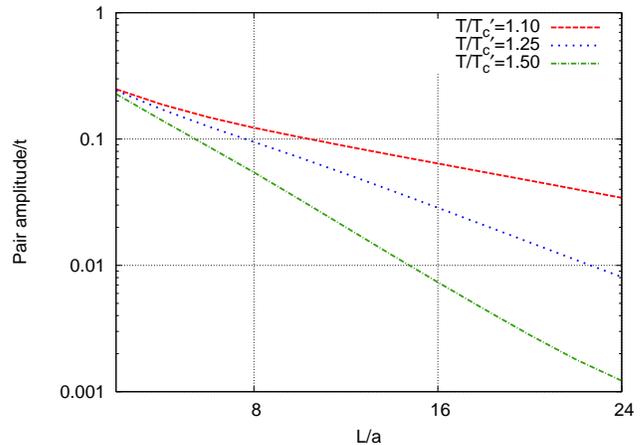,height=\columnwidth}
\end{turn}
\caption{(Color online) The pair amplitude at L/2 as a function of L for different
temperatures above $T^\prime_c$ for the SN$^\prime$S system with $U^\prime=-2t$ and $U=-4t$.}
\end{center}
\end{figure}

\begin{figure}[ht]
\begin{center}
\begin{turn}{-90}
\epsfig{figure=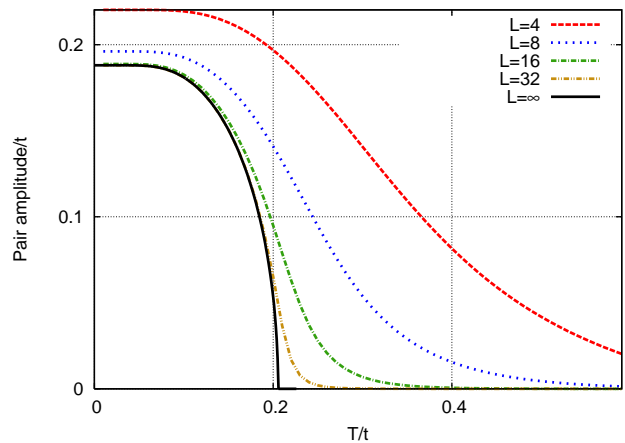,height=\columnwidth}
\end{turn}
\caption{(Color online) The pair amplitude at L/2 as a function of temperature for
different L for the SN$^\prime$S system with $U^\prime=-2t$ and $U=-4t$.}
\end{center}
\end{figure}
If we plot the order parameter as a function of temperature for
different lengths (L) of the weak link (Fig. 3), we can observe two
main effects. First, at $T=0$ K, the proximity effect will modify
the order parameter at $L/2$. For lengths smaller than 10 lattice
constants this effect is important. Because the $N^\prime$ layer is
superconducting at $T=0$K, the main length scale in this layer is
the superconducting coherence length $\xi=\hbar v_F /\Delta$. The
second effect is observed at temperatures close to $T^\prime_c$. If
the $N^\prime$ layer was not connected to the superconducting
layers, then, according to the mean-field behavior, the order
parameter would vanish at $T^\prime_c$. For temperatures higher than
$T^\prime_c$ the $N^\prime$ layer cannot sustain superconductivity
by itself. It is only in the presence of the $S$ layers, that the
order parameter at $L/2$ has non-zero values. Note that {\em the
length $L$ for which we obtain non-zero values of the order
parameter above $T_c^\prime$ is much larger than the value of the
length beyond which effects are unobservable at $T=0$ K}. In Fig. 4,
we compare the order parameters at $L/2$ for two cases: $U^\prime=0$
and $U^\prime=-2t$. We observe that the order parameter for the case
$U^\prime=-2t$ is larger and that close to $T^\prime_c$ the
discrepancy is enhanced. This is a clear indication that the Cooper
pair leaking distance is larger in the case of a non-zero
$T^\prime_c$.

\begin{figure}[ht]
\begin{center}
\begin{turn}{-90}
\epsfig{figure=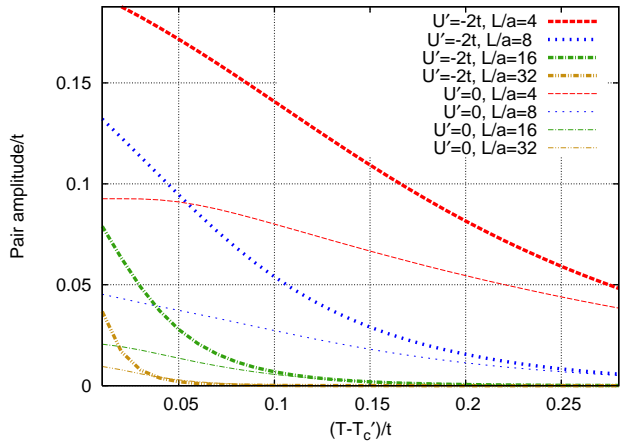,height=\columnwidth}
\end{turn}
\caption{(Color online) The pair amplitude at L/2 as a function of relative temperature for different L for the SN$^\prime$S system for $U=-4t$, $U^\prime=0$ and $U=-4t$, $U^\prime=-2t$.}
\end{center}
\end{figure}

\begin{figure}[ht]
\begin{center}
\begin{turn}{-90}
\epsfig{figure=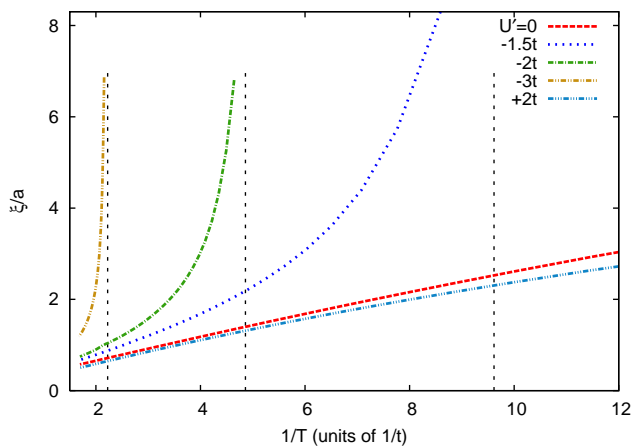,height=\columnwidth}
\end{turn}
\caption{(Color online) The leaking distance as a function of 1/T for different
interaction parameters $U^\prime$ for the SN$^\prime$S system with $U=-4t$. The vertical dashed lines represent the inverse of the critical temperatures for the corresponding $U^\prime$ parameters: $T_c(U^\prime=-1.5t)=0.104t$, $T_c(U^\prime=-2t)=0.205t$, $T_c(U^\prime=-3t)=0.46t$. }
\end{center}
\end{figure}

In order to investigate further the dependence of the leaking
distance on the magnitude of the superconducting correlations in the
$N^\prime$ layer, in Fig. 5, we summarize the extracted leaking distance
obtained for different parameters. The $T_c^\prime = 0$K line(dashed) is inversely
proportional to the temperature, as expected. For $T_c^\prime > 0$ K the
leaking distance is diverging at $T_c^\prime$; this, of course leads to a
giant proximity effect near these temperatures, as the figure visually
demonstrates. Another feature of the calculation
is that for any given temperature, a higher $T_c^\prime$ will result in a
larger leaking distance. For repulsive on-site interactions,
$U^\prime=+2t$, the leaking distance is even smaller than in the normal
metal case. This is a clear demonstration of the fact that
interactions in the $N^\prime$ layer will influence the way Cooper pairs
leak from the superconducting side. Such ``feedback" will not be captured
in calculations that are not self-consistent.

The proximity effect can be observed either by growing
superconducting thin films on top of normal metals and measuring the
critical temperature of the system, or by forming a Josephson
junction and measuring the Josephson current through the weak link.
If the two superconducting sides are not coupled then there is no
Josephson current. As we bring the superconducting sides closer to
one another, the
proximity effect will influence the value of the order parameter in
the $N^\prime$ layer. A non-zero value of the order parameter throughout
the whole system will result in a non-zero value of the Josephson
current.

The BdG equations are well suited for calculating the dc Josephson
current. In order to have current between the two superconducting
sides, the order parameters in the two sides have to have different
phases.

\begin{figure}[ht]
\begin{center}
\begin{turn}{-90}
\epsfig{figure=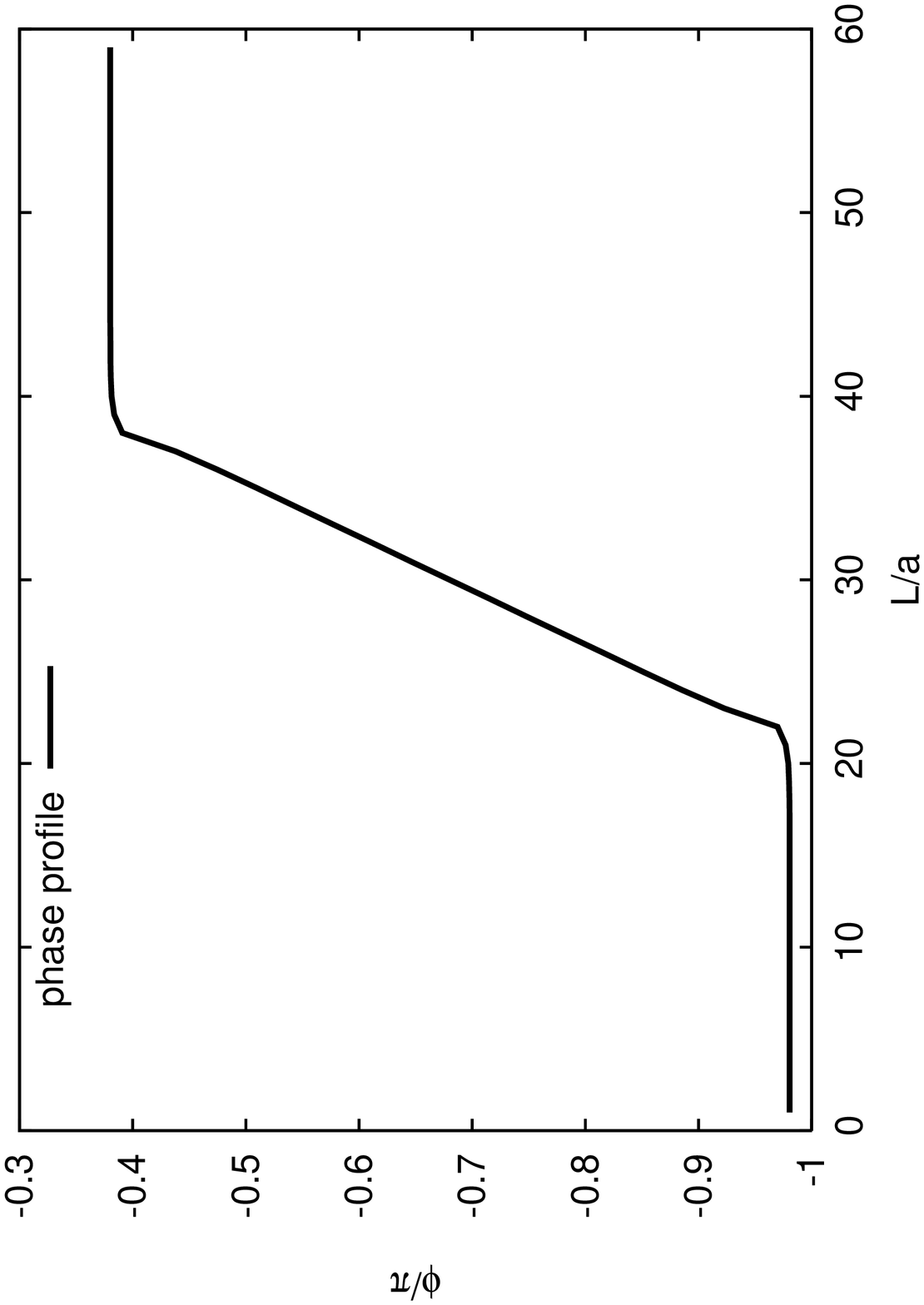,height=0.7\columnwidth}
\epsfig{figure=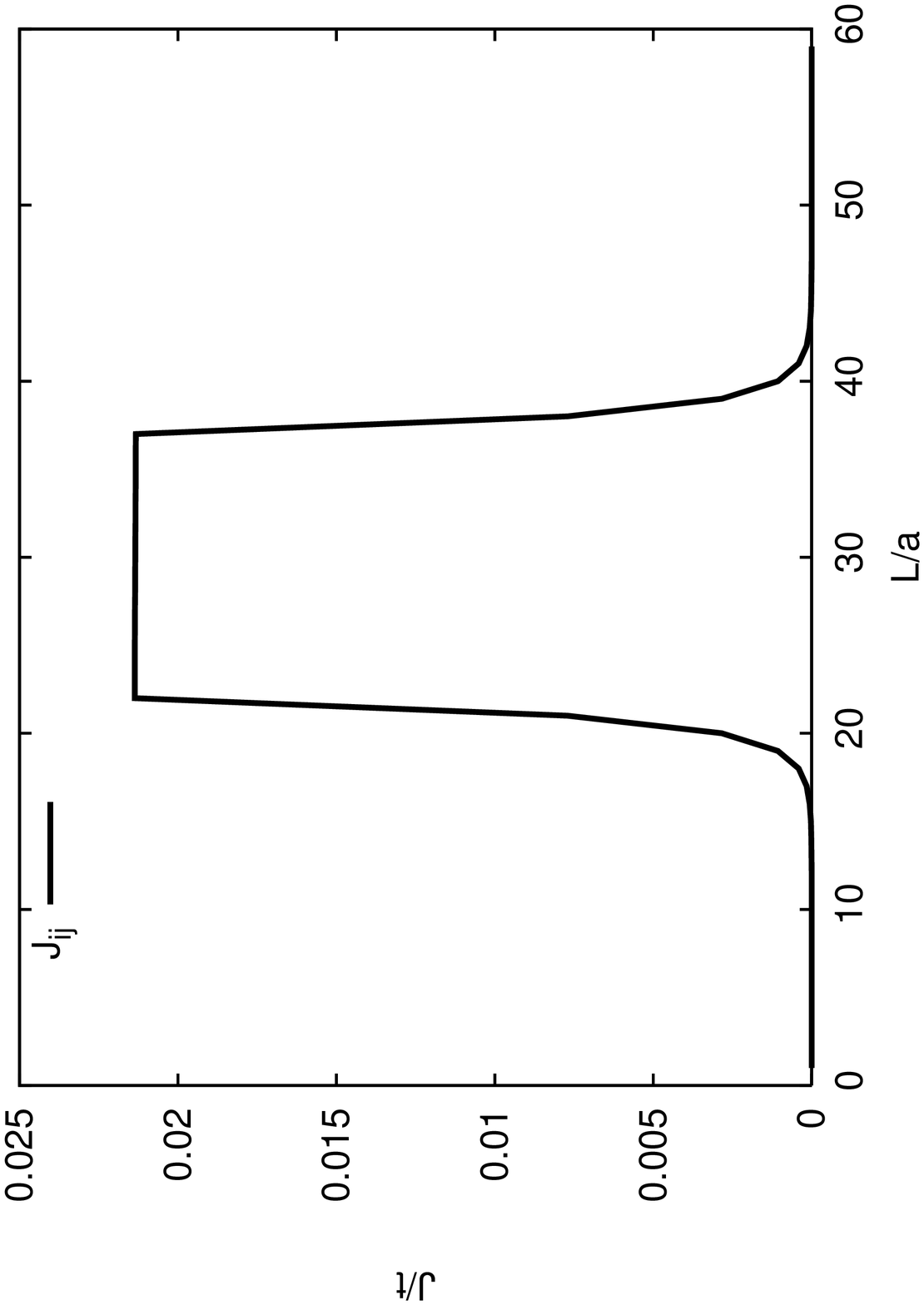,height=0.7\columnwidth}
\end{turn}
\caption{(a) Phase profile and (b) dc Josephson current as a function of position. The phase is calculated self-consistently only in the middle layer and the continuity equation is satisfied only in this layer.}
\end{center}
\end{figure}

In our calculation we fix the phases of the
order parameter on the $S$ layers, and our self-consistent
calculation will give the magnitude and the phase of the order
parameter in the $N^\prime$ side. The results of such a calculation are
shown in Fig. 6a and 6b. The phase of the order parameter in $N^\prime$
will vary continuously from $\phi_{left}$ to $\phi_{right}$ and the
dc Josephson current will be constant throughout the layer. An
interesting case is the one where $\Delta \phi = \pi$, for which
there is a phase-slip point at $L/2$. Right at the phase-slip the
order parameter vanishes. In order to extract only the proximity
effect from the current calculation, we need to find the phase
difference for which the current is maximal. For a point contact
Josephson junction the current has the following behavior
\cite{ambegaokar}:
\begin{equation}
J=J_m sin(\Delta \phi),
\end{equation}
while for a long junction it deviates from the sinusoidal behavior
\cite{rodero}.

\begin{figure}[ht]
\begin{center}
\begin{turn}{-90}
\epsfig{figure=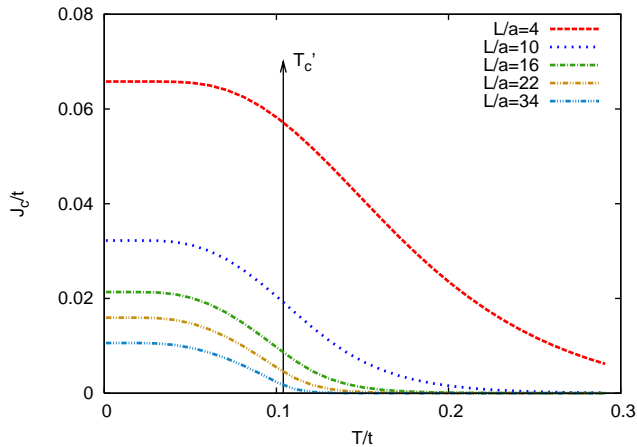,height=\columnwidth}
\end{turn}
\caption{(Color online) The dc Josephson current in the middle layer as a function of temperature for
different lengths L of the weak link for the SN$^\prime$S system with $U=-3t$ and $U^\prime=-1.5t$. The arrow represents the critical temperature of the middle layer, $T_c(U^\prime=-1.5t)=0.104t$.}
\end{center}
\end{figure}

We calculate the dc Josephson current for different lengths of the
weak link and for different temperatures for $U^\prime=-1.5t$ and $U=-3t$.
The results are summarized in Fig. 7. When the two superconducting
layers (S) are close together the proximity effect modifies the
magnitude of the order parameter at L/2 in the N$^\prime$ layer. A large
value of the order parameter will give a large value for the
current. As L increases the order parameter decreases exponentially.
This results in a decay of the current as a function of L.
The main result is that the behavior of the
Josephson current as a function of L and T reflects the existence of
a leaking distance larger than the one expected from a
normal metal. For $U^\prime=-1.5t$ and $T=0.125t$, the normal metal gives a leaking
distance of $\xi_0\ \sim 2a$ while the self-consistently calculated
one gives a value of $\xi \sim 7a$. This is seen in Fig. 7, where for
$L=16a$ the current is non-zero for temperatures close to but greater
than $T_c^\prime$, and it has a linear dependence on temperature near
$T_c^\prime$.

As shown in previous attempts to explain the ``giant proximity
effect", the presence of pockets of superconductivity in the
N$^\prime$ layer will greatly enhance the current through the
system, even for long weak links. Coupled with the enhancement of
the leaking distance around $T_c^\prime$ the presence of the
superconducting pockets will effectively decrease the length of the
weak link. We consider equally spaced superconducting areas with
on-site interactions of strength $U=-4t$ embedded in the weak link with
interaction strength $U=-2t$. In Fig. 8 we show the Josephson
current for different lengths of the weak link and with
superconducting pockets occupying a volume percentage $p=0.2$ of the
weak link. The size of the considered pockets is one lattice site.
The effect on the Josephson current is drastic --- the current has
non-zero values well above $T_c^\prime$. We also notice that for
this volume of embedded superconductivity, the current has a weak
dependence on the length of the junction.

\begin{figure}[ht]
\begin{center}
\begin{turn}{-90}
\epsfig{figure=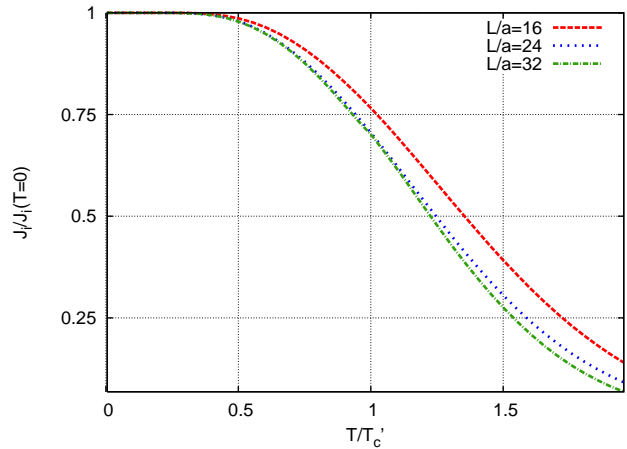,height=\columnwidth}
\end{turn}
\caption{(Color online) The dc Josephson current in the middle layer as a function of temperature for
different lengths L of the weak link for the SN$^\prime$S system with $U=-4t$ and $U^\prime=-2t$. Equally spaced areas of superconductivity in the N$^\prime$ layer are considered. The percent volume of the pockets of superconductivity with $U=-4t$ is $p=0.2$.}
\end{center}
\end{figure}

\begin{figure}[ht]
\begin{center}
\begin{turn}{-90}
\epsfig{figure=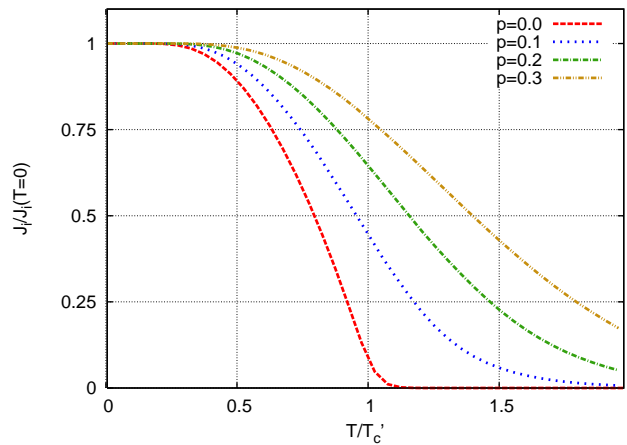,height=\columnwidth}
\end{turn}
\caption{(Color online) The dc Josephson current in the middle layer as a function of temperature for different percent volumes of embedded superconductivity in N$^\prime$ for the SN$^\prime$S system with $U=-4t$ and $U^\prime=-2t$. The length L of the weak link is $L=40a$.}
\end{center}
\end{figure}

The strength of the coupling between the exterior superconductors
will be given by the volume of these pockets. This is seen in Fig.
9, where we plot the Josephson current for a weak link of length
$L=40a$ as a function temperature for different percent volumes of
strong superconducting pockets embedded in the weak superconductor.
As expected, for $p=0.0$ the junction is too long to couple the
strong superconductors and the current vanishes above but very close
to $T_c^\prime$. Increasing $p$, the junction will effectively
shorten and the two exterior superconductors will be coupled well
above $T_c^\prime$.

\section{Results for DN$^\prime$D}

The d-wave symmetry of the order parameter is attained if we
consider nearest-neighbor interactions $V_{i i+\delta}=-V_i$ and
vanishing or repulsive on-site interactions. The d-wave order
parameter is calculated in the following way:
\begin{equation}
\Delta_d(i)=\frac{1}{4}(\Delta_x(i)+\Delta_{-x}(i)-\Delta_{y}(i)-\Delta_{-y}(i)),
\end{equation}
where $\Delta_{x}$ describes superconducting correlations in the
$\hat{x}$ direction. In a similar manner, as detailed in the SN$^\prime$S
case, we set up $V_i$ so that $V_i=V$ for $0<i<A$ and $B<i<N$, while
$V_i=V^\prime$ for $A<i<B$. This will allow us to describe a weak link
with a non-zero critical temperature.

\begin{figure}[ht]
\begin{center}
\begin{turn}{-90}
\epsfig{figure=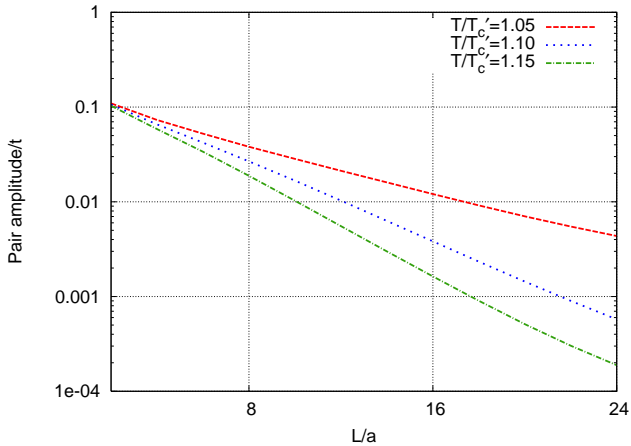,height=\columnwidth}
\end{turn}
\caption{(Color online) D-wave order parameter at L/2 as a function L for different
temperatures - 100 d-wave case with $V=-4t$ and $V^\prime=-2t$.}
\end{center}
\end{figure}

\begin{figure}[tp]
\begin{center}
\begin{turn}{-90}
\epsfig{figure=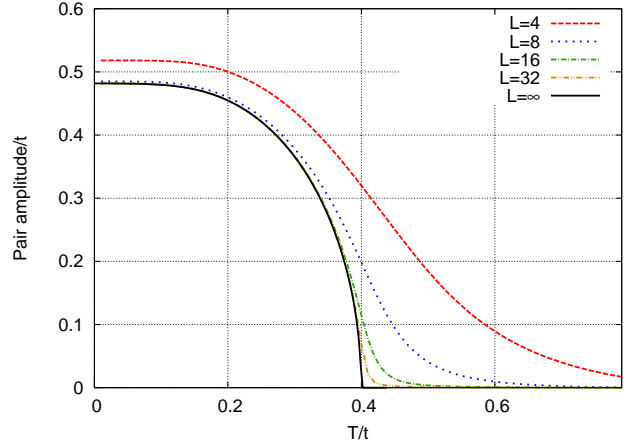,height=\columnwidth}
\end{turn}
\caption{(Color online) D-wave order parameter at L/2 as a function of T for different
lengths - 100 d-wave case with $V=-4t$ and $V^\prime=-2t$.}
\end{center}
\end{figure}

\begin{figure}[tp]
\begin{center}
\begin{turn}{-90}
\epsfig{figure=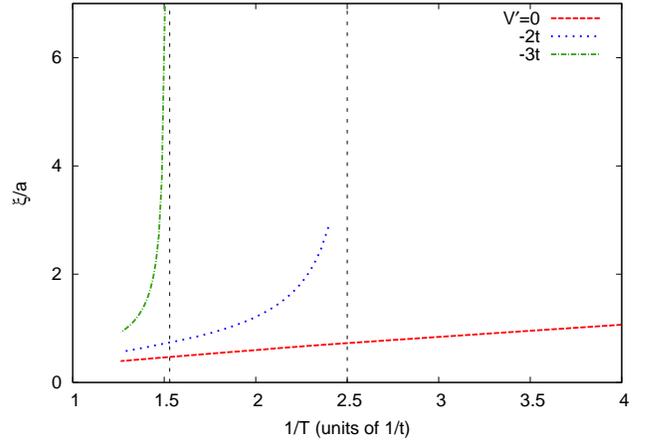,height=\columnwidth}
\end{turn}
\caption{(Color online) Leaking distance as a function of inverse temperature for
different interaction strengths - 100 d-wave case with $V=-4t$. The vertical dashed lines represent the inverse of the critical temperatures for the corresponding $V^\prime$ parameters: $T_c(V^\prime=-2t)=0.4t$, $T_c(V^\prime=-3t)=0.67t$.}
\end{center}
\end{figure}

For the 100 interface (between the $a-b$ planes of a high-$T_c$
superconductor), the dependence of the order parameter is very
similar to the s-wave case. Fig. 10 shows the semi-log plot of the
order parameter as a function of distance from the interface for
different temperatures. Again, we can observe the exponential decay
and define the leaking distance $\xi$. The dependence of the order
parameter on temperature for different lengths of the weak link is
shown in Fig. 11 and the two manifestations of the proximity effect
are seen. First at $T=0$K the order parameter is modified if $L/2$
is of the order of the superconducting coherence length in the N$^\prime$
layer. Secondly, above $T^\prime_c$ the order parameter decays with
increasing temperature but has non-zero value even if $L/2$ is
greater than the conventional leaking distance defined by the
$T^\prime_c=0$K case. The self-consistently calculated leaking distance is
shown in Fig. 12. Similar to the s-wave case it diverges at $T^\prime_c$
and, for the same temperature, larger interactions in the weak
superconductor will increase the leaking distance.

\begin{figure}[ht]
\begin{center}
\begin{turn}{-90}
\epsfig{figure=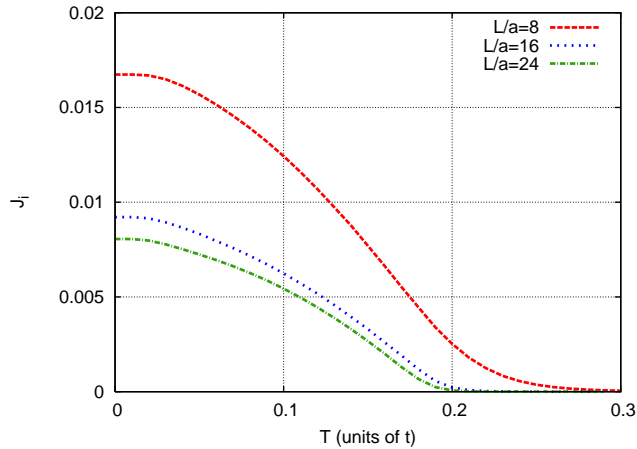,height=\columnwidth}
\end{turn}
\caption{(Color online) The c-axis dc Josephson current in the middle layer as a function of temperature for
different lengths L of the weak link for which $V=-4t$ and $V^\prime=-2t$. The c-axis hopping amplitude is $t^\perp=0.5t^{||}$.}
\end{center}
\end{figure}

The coherent transport in the c-axis direction will be described by
the hopping amplitude in the $\hat{z}$ direction,
$t^{\perp}=0.5t^{||}$. Fig. 13 shows the Josephson critical current
as a function of temperature for different lengths of the weak
superconducting layer. The general behavior is similar to the one of
the s-wave junction. The d-wave order parameter has no $k_z$
dependence, and thus in the $\hat{z}$ direction it is an effective
on-site order parameter. For short weak links the current does not
vanish abruptly above $T_c^\prime$ but rather has a smooth
dependence on temperature. This dependence on temperature above
$T_c^\prime$ shows that the length of the weak link is comparable to
the leaking distance in this layer. The increase of the leaking
distance due to the finite $T_c^\prime$ cannot explain by itself the
observed ``giant proximity effect". It is only the conjunction with
the presence of disordered pockets of superconductivity in the weak
link that makes this effect possible. The calculation of the
Josephson current in the presence of the disordered pockets from the
previous section stands also for the c-axis geometry. The extra
dimension will only affect the necessary volume of superconductivity
needed to observe a ``giant proximity effect".

\section{Summary}

In summary, using a tight binding formulation of the extended
Hubbard Hamiltonian, we solve the BdG equations for a system
composed of three layers: two superconducting layers (either s-wave or
d-wave), and a weaker superconductor sandwiched in between.
We examined the proximity effect
induced by the exterior superconducting layers in the ``normal metal" interior layer.
We observed that, in agreement with previous calculations, the order
parameter has an exponential decay behavior, the characteristic
decay length being the leaking distance. In both s-wave and d-wave
cases the leaking distance is only dependent on the properties of
the N$^\prime$ layer.
For $T^\prime_c=0$K (normal metal) for both s-wave and d-wave
symmetries the leaking distance is inversely proportional to the
temperature. If $T^\prime_c>0$K, the leaking distance diverges at $T^\prime_c$
and at the same temperature larger attractive interactions in the
middle layer will
increase the leaking distance. Essentially, the BdG formalism provides a means
for the normal layer to feel pairing fluctuations above its critical temperature,
$T_c^\prime$. These are not spontaneous, in that they arise from an `applied'
pairing field produced by the outer layers. This accounts for the much higher
leaking distance for a weak superconductor.

We also calculated the dc Josephson
current, and extracted the maximum value. We observed that the
current has a non-zero value for lengths of the weak link much
larger than the ``conventional" leaking distance, and for
temperatures well above $T^\prime_c$. Although the divergence of the leaking
distance at the critical temperature of the N$^\prime$ layer enhances the Josephson current for temperatures above $T_c^\prime$, it is not enough to explain the experimental measurement of the ``giant proximity effect" \cite{bozovic}. As prompted by previous attempts to explain the ``giant proximity effect" \cite{kresin,dagotto}, we considered areas of superconductivity with critical temperature $T_c>T_c^\prime$, which are embedded in the N$^\prime$ layer. Further enhancement of the Josephson current is observed. Depending on the volume of the superconducting pockets, non-zero values of the Josephson current are obtained even for temperatures $T>2T_c^\prime$. These results form the basis for a qualitative understanding of the giant proximity effect observed by Bozovic et al. \cite{bozovic}

\begin{acknowledgments}
We thank Wonkee Kim and Fatih Do\u gan for useful discussions. This work was supported in part by the Natural Sciences and Engineering Research Council of Canada (NSERC), by ICORE (Alberta), and by the
Canadian Institute for Advanced Research (CIAR).
\end{acknowledgments}

\bibliographystyle{prl}

\begin{thebibliography}{1}

\bibitem{bozovic} I. Bozovic et al., \prl {\bf 93}, 157002 (2004).
\bibitem{ginzburg} V. L. Ginzburg and L. D. Landau, Zh. Eksp. Theor. Fiz. \textbf{20}, 1064 (1950).
\bibitem{chen} Jian Hua Chen, \prb {\bf 42}, 3952 (1990); \prb {\bf 42}, 3957 (1990).
\bibitem{kogan} V. G. Kogan, \prb \textbf{26}, 88 (1982).
\bibitem{gennes64} P. G. de Gennes, Rev. Mod. Phys. \textbf{36}, 225 (1964).
\bibitem{werthamer} N. R. Werthamer, Phys. Rev. \textbf{132}, 2440 (1963).
\bibitem{macmillan} W. L. McMillan, Phys. Rev. \textbf{175}, 559 (1968); Phys. Rev. \textbf{175}, 537 (1968).
\bibitem{wu} J.Z. Wu, X.X. Yao, C.S Ting, W.K. Chu, \prb \textbf{46}, 14059 (1992).
\bibitem{hirsch} J. E. Hirsch,Physica C \textbf{194}, 119 (1992).
\bibitem{zhu} J.-X. Zhu and C. S. Ting, \prb \textbf{61}, 1456 (2000).
\bibitem{halterman} K. Halterman and O. T. Valls, \prb \textbf{65}, 014509 (2002).
\bibitem{likharev} K. K. Likharev, Rev. Mod. Phys. {\bf 51}, 101 (1979).
\bibitem{golubov} A.A. Golubov et al., Rev. Mod. Phys. {\bf 76}, 411 (2004).
\bibitem{tanaka1} Y. Tanaka, S. Kashiwaya, \prl {\bf 74}, 3451 (1995).
\bibitem{tanaka2} Y. Tanaka, S. Kashiwaya, \prb {\bf 56}, 892 (1997).
\bibitem{tanaka3} Y. Tanaka, S. Kashiwaya, \prb {\bf 53}, R11957 (1997).
\bibitem{tanaka4} Y. Tanaka, Yu. V. Nazarov, A. A. Golubov, S. Kashiwaya, \prb {\bf 69}, 144519 (2004).
\bibitem{tanaka5} Y. Tanaka, Yu. V. Nazarov, S. Kashiwaya, \prl {\bf 90}, 167003-1 (2003).
\bibitem{delin} K. A. Delin, A. W. Kleinsasser, Supercond. Sci. Technol. {\bf 9} (1996) 227.
\bibitem{usadel} K. D. Usadel, \prl {\bf 25}, 507 (1970).
\bibitem{cuevas05} For a more recent application of the Usadel formalism, see,
for example, J.C. Cuevas, J. Hammer, J. Kopu, J.K. Viljas, and M.
Eschrig, cond-mat/0507247.
\bibitem{kresin} V. Kresin, Yu. Ovchinnikov, S. Wolf, Appl. Phys. Lett. {\bf 83} 722 (2003).
\bibitem{dagotto} G. Alvarez, M. Mayr, A. Moreo, E. Dagotto, \prb {\bf 71}, 014514 (2005).
\bibitem{zhu05} J.-X. Zhu, cond-mat/0508646.
\bibitem{mcelroy} K. McElroy \textit{et al.}, Science {\bf 309}, 1048 (2005).
\bibitem{micnas} R. Micnas, J. Ranninger and S. Robaszkiewicz,  Rev. Mod. Phys. {\bf 62}, 113 (1990).
\bibitem{gennes} P.G. de Gennes, \textit{Superconductivity of Metals and Alloys} (Benjamin, New York, 1966).
\bibitem{ambegaokar} V. Ambegaokar and A. Baratoff, \prl {\bf 10},
486 (1963).
\bibitem{rodero} A. Mart\'in-Rodero, F. J. Garc\'ia-Vidal, and A. L.
Yeyati \prl {\bf 72}, 554 (1994).
\end{thebibliography}

\end{document}